\documentclass[aps,pra,twocolumn,showpacs]{revtex4-1}

\usepackage{indentfirst}
\usepackage{bm}
\usepackage{verbatim}
\usepackage{graphicx}
\usepackage{subfigure}
\usepackage{epsfig}
\usepackage{amsmath}
\usepackage{booktabs}
\usepackage{multirow}
\usepackage{txfonts}
\usepackage{float}
\usepackage{tabularx}
\usepackage{threeparttable}

\usepackage[citecolor=blue,colorlinks=true,linkcolor=blue,urlcolor=blue,breaklinks=true]{hyperref}

\begin{document}

\title{Long-time signatures of chaos in large atom-light frequency ratios Rabi model}

\author{Shangyun Wang$^{1}$}
\author{Songbai Chen$^{2,3}$}
\author{Jiliang Jing$^{2,3}$}

\email[E-mail address: ]{sywang@hynu.edu.cn}
\affiliation{$^{1}$College of Physics and Electronic Engineering, Hengyang Normal University, Hengyang 421002, China \\
$^{2}$Key Laboratory of Low-Dimensional Quantum Structures and Quantum Control of Ministry of Education, Key Laboratory for Matter Microstructure and Function of Hunan Province, Department of Physics and Synergetic Innovation Center for Quantum Effects and Applications, Hunan Normal University, Changsha 410081, China \\
$^{3}$Center for Gravitation and Cosmology, College of Physical Science and Technology, Yangzhou University, Yangzhou 225009, People's Republic of China}

\begin{abstract}
As one of the famous effects in quantum Rabi model (QRM), Rabi oscillation may lead to the occurrence of quantum dynamics behaviors without classical dynamic counterparts, such as quantum collapse and revival effects.
In this paper, we focus on studying the long-time quantum signatures of chaos in the large atom-light frequency ratios Rabi model.
It is shown that the saturated values of the entanglement entropy for initial states located in chaotic sea are higher than that in the regular regions, and the Husimi Q function are more dispersed in phase space.
Moreover, we observed that the long-time average entanglement entropy and spin variance correspond well with the semiclassical phase space.
Our results imply that the correspondence principle does not invalidated by quantum collapse and revival effects in the large atom-light frequency ratios Rabi model.


\end{abstract}
\maketitle

\section{Introduction}
Correspondence principle is an indispensable cornerstone of quantum mechanics, which bridges classical and quantum world. In classical physics, the hallmark of chaotic dynamics is the extreme sensitivity of the time evolution of a system to initial conditions. However, quantum mechanics does not support a similar definition due to the uncertainty principle and the overlap between quantum states.
This fundamental incompatibility poses a challenge to the correspondence principle and has motivated a long-standing search for chaotic signatures in quantum systems~\cite{csqs1,csqs2,csqs3,csqs4,csqs5}.

Defining quantum chaos by analogy with classical chaos faces numerous obstacles.
Finding quantum tools which exhibit significant differences between chaotic and regular regions in semiclassical systems has become one of the mainstream solutions for studying quantum chaos. The study of entanglement entropy indicates that in semiclassical quantum systems, the saturated values of entanglement entropy in the chaotic regions are significantly greater than that in the regular regions~\cite{EE1,EE2,EE3,EE4,EE5,EE6,EE7,EE8,EE9}. Specifically, the corresponding relationship between the time-averaged entanglement entropy and the classical phase space is one of the important evidences to test the correspondence principle~\cite{EECP1,EECP2}.
Because chaos is inherently a dynamical phenomenon, as one of the quasi-probability distribution functions used to visualize quantum wave packets dynamics, the Husimi Q function exhibits significant time-evolution differences between chaotic and regular regions~\cite{EECP1,EECP2,Husimi1,Husimi2,Husimi3,Husimi4,Husimi5,Husimi6}.
In the classical spin limit, the spin variances of chaotic sea and regular regions exhibit significant differences~\cite{sl1}.
Moreover, the level spacing distribution~\cite{lsd1,lsd2,lsd3,lsd4,lsd5,lsd6}, Loschmidit Echo~\cite{heo1,heo2,heo3,heo4,heo5,heo6} and out-of-time-
ordered correlator (OTOC)~\cite{ot1,ot2,ot3,ot4,ot5,ot6,ot7,ot8,ot9,ot10,ot11,ot12,ot13} have been proven to be effective tools for exploring chaos in quantum systems.

For the systems of light interacting with atoms, the researches related to quantum chaos mainly focus on the Dicke model~\cite{Dicke1,Dicke3,Dicke4,Dicke5,Dicke6,Dicke7,Dicke8,Dicke9}. The Rabi model~\cite{Rabi}, as the simplest version of the Dicke model, describes a single mode cavity field and a two-level atom which interacting via dipolar coupling~\cite{Rabi2,Rabi3}. Since the Rabi model exhibits similar interaction mechanism to Dicke model, the chaotic dynamic behaviors should appear in this system~\cite{Rabi4}. However, the study of chaos in this model has been neglected for a long time owing in part to the number of atom, and in part to the Rabi oscillations, which generate some effects without classical dynamic counterparts such as quantum collapse and revival~\cite{husimicrthe1}.

The quantum chaos of a system with quantum collapse and revival effect is of great interest, not only from the horizon of quantum dynamics, but from the viewpoint of the correspondence between quantum and classical mechanics as well. Recently, Irish and Armour~\cite{scRabi} studied the semiclassical limit in the QRM and their result provides a important and feasible framework to study the classical-quantum correspondence in this system.
Kirkova \emph{et al.}~\cite{scRabi2} found that the OTOC quickly saturates in the normal phase and undergoes exponential growth in the superradiant phase of QRM when the ratio of level-splitting $\omega$ to bosonic frequency $\omega_0$ grows to infinity $\eta=\omega/\omega_0\rightarrow \infty$.
This indicates that chaotic signatures appear in the QRM at effective thermodynamic limits $\eta\rightarrow \infty$. On the other hand, using the mean field approximation theory, Ref.~\cite{scRabi3} shows that quantum collapse and revival effects can interfere with the exponential behavior of OTOC in the anisotropic quantum Rabi model.
However, it is not yet uncertain that whether the quantum collapse and revival effects affect the long-time signatures of entanglement entropy and Husimi Q function, which serve as tools for diagnosing quantum chaos. In addition, since there is only one atom in the Rabi model, the difference of spin variance values between the chaotic and regular regions is still unclear and needs to be investigated further.

In this paper, we will focus on studying the long-time quantum signatures of chaos in QRM which consists of a single cavity field mode interact with a two-level atom via dipolar coupling. We find that both the long-time average entanglement entropy and the average spin variance correspond well with the semiclassical phase space structure. Moreover, the Husimi Q function exhibits significant differences between the chaotic and regular regions while the splitting and merging behaviors emerge in its evolutionary process.

\begin{figure}[ht]
\begin{center}
\includegraphics[width=6cm]{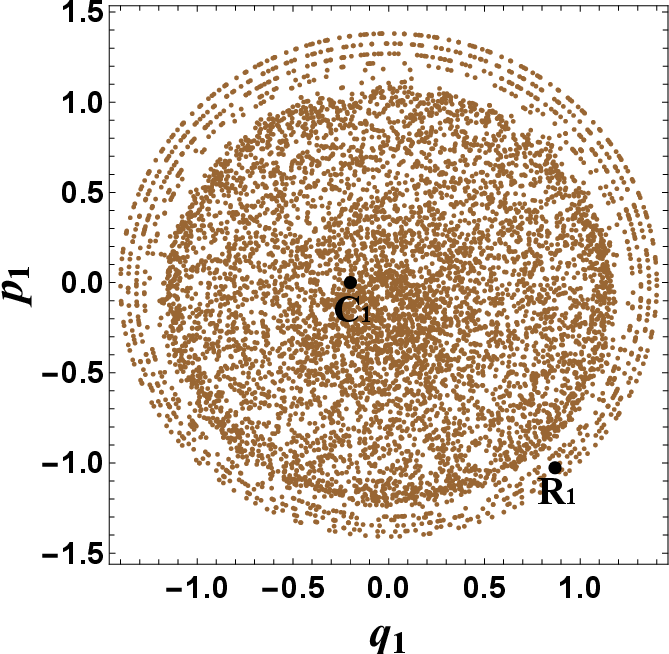}
\caption{The Poincar\'{e} section for the quantum Rabi model in the case: $q_{2}=0, p_{2}>0$, with $\omega=18$, $\omega_{0} =1$, $g=4$ and the system energy $E = 14$. Point $R_1 (q_{1} = 0.86853,p_{1} =-1.02681,q_{2} =0,p_{2} =3.66657)$ and $C_1 (q_{1} = -0.2,p_{1} =0,q_{2} =0,p_{2} =6.72904)$ are situated in stable island and chaotic sea respectively.}\label{f1}
\end{center}
\end{figure}
This paper is organized as follows: In Sec. \ref{sec2}, we introduce briefly the large atom-light frequency ratios Rabi model and its semiclassical phase space. In Sec. \ref{sec3}, we investigate the truncated photon number in the QRM when the ratio of level-splitting $\omega$ to bosonic frequency $\omega_0$ grows to $\eta=\omega/\omega_0\rightarrow 18$. In Sec. \ref{sec4}, we study the correspondence between the distribution of long-time averaged entanglement entropy and the classical Poincar\'{e} section.
In Sec. \ref{sec5}, we analyze the evolutionary differences of Husimi Q function with splitting and merging behaviors between chaotic and regular regions in this model.
In Sec. \ref{sec6}, we discuss the correspondence between the distribution of time-averaged spin variance and the classical Poincar\'{e} section.
Finally, we present results and a brief summary.

\section{QUANTUM RABI MODEL}\label{sec2}
Let us now briefly introduce the quantum Rabi model which is one of the simplest and
most fundamental models describing quantum light-matter interaction. The Rabi
Hamiltonian can be expressed as ($\hbar=1$)
\begin{eqnarray}
\hat{H}&=& \frac{\omega}{2} \hat{\sigma}_{z} + \omega_0 \hat{a}^{\dagger}\hat{a} + g(\hat{a}^{\dagger} + \hat{a})(\hat{\sigma}_{+}+ \hat{\sigma}_{-}), \label{ha1}
\end{eqnarray}
where $\omega$ is the level-splitting of two-level atom,
$\hat{a}^{\dagger}$ and $\hat{a}$ are respectively the creation and annihilation operators of the single-mode cavity with frequency $\omega_0$. The coupling $g$ is the strength of the dipolar atom-field interaction and $\hat{\sigma}_{+}$,  $\hat{\sigma}_{-}$ are atomic raising and lowing operators, respectively.

\begin{figure}[ht]
\begin{center}
\includegraphics[width=6cm]{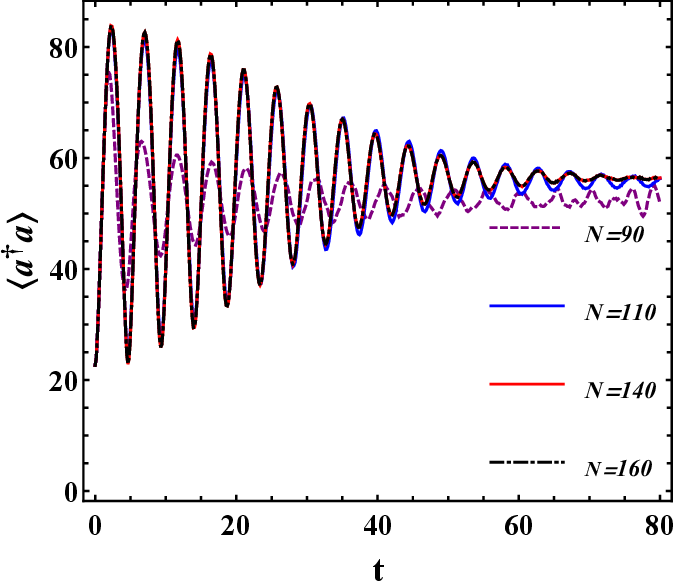}
\caption{ Time evolution of the average photon number $\langle \hat{a}^{\dagger}\hat{a}\rangle$ with
different the initial system photon number $N$ for the initial coherent states
centered at chaotic point $C_1$.}\label{f2}
\end{center}
\end{figure}
To study the quantum signatures of chaos in the quantum Rabi model. As in Refs.~\cite{EE1}, we take the initial states to be coherent states since the its uncertainty in phase space are the smallest. The initial quantum states are chosen as
\begin{eqnarray}
|\psi(0)\rangle &=& |\tau\rangle\otimes|\beta\rangle,\label{init1}
\end{eqnarray}
with
\begin{eqnarray}
|\tau\rangle &=& (1+\tau\tau^{*})^{-\frac{1}{2}}\mathrm{e}^{\hat{\tau} \sigma_{+}}|\frac{1}{2},-\frac{1}{2}\rangle,\label{acoh1}\\
|\beta\rangle &=& \mathrm{e}^{-\beta\beta^{*}/2}\mathrm{e}^{\beta \hat{a}^{\dagger}}|0\rangle, \label{bcoh1}
\end{eqnarray}
and
\begin{eqnarray}
\tau &=& \frac{q_1 + \mathrm{i} p_1}{\sqrt{2-q^{2}_1 -p^{2}_1}},  \ \ \ \ \ \beta = (q_2 + \mathrm{i}p_2)/\sqrt{2},   \label{tb1}
\end{eqnarray}
where $|\tau\rangle$ and $|\beta\rangle$ are Bloch coherent states of atom and Glauber coherent states of bosons, and the states $|\frac{1}{2},-\frac{1}{2}\rangle$ and $|0\rangle$ are the ground state of two-level atom and the vacuum state of single-mode cavity field, respectively. With the mean field approximation procedure~\cite{mfap},
\begin{eqnarray}
\langle\tau|\hat{\sigma}_{+}|\tau\rangle &=& \frac{\tau^*}{1+\tau\tau^*},
\end{eqnarray}
\begin{eqnarray}
\langle\tau|\hat{\sigma}_{-}|\tau\rangle &=& \frac{\tau}{1+\tau\tau^*},
\end{eqnarray}
\begin{eqnarray}
\langle\tau|\hat{\sigma}_{z}|\tau\rangle &=& -\frac{1-\tau\tau^*}{1+\tau\tau^*},
\end{eqnarray}
\begin{eqnarray}
\langle\beta|\hat{a}|\beta\rangle &=& \beta,
\end{eqnarray}
\begin{eqnarray}
\langle\beta|\hat{a}^{\dagger}|\beta\rangle &=& \beta^*,
\end{eqnarray}
the semiclassical Rabi Hamiltonian reads
\begin{eqnarray}
H_{cl} &\equiv& \langle \tau\beta|\hat{H}|\tau\beta\rangle = \frac{\omega}{2} (q^{2}_1 +p^{2}_1 -1)+ \frac{\omega_0}{2}(q^{2}_2 + p^{2}_2) \notag \\
&+& g q_1 q_2\sqrt{4-2(q^{2}_1 +p^{2}_1)},\label{ha2}
\end{eqnarray}
where $q_2=(a^{\dag} + a)/\sqrt{2}$ and $p_2= \mathrm{i}(a^{\dag} - a)/\sqrt{2}$.
With the classical Hamiltonian (\ref{ha2}), we obtain the poicar\'{e} section
of the QRM, as shown in Fig. \ref{f1}. This mixed phase space section contains both stable island and chaotic sea composed of many discrete points. Motion across the boundaries between regular and chaotic regions is classically forbidden.

\section{TRUNCATED PHOTON NUMBER}\label{sec3}

QRM is intuitively far from the so-called thermodynamic limit ($N\rightarrow\infty$) because there is only one atom, and it seems impossible to achieve the classical-quantum correspondence in this system.
On the other hand, the exponential growth of OTOC in the superradiant phase of QRM when the ratio of level-splitting $\omega$ to bosonic frequency grows to infinity~\cite{scRabi2}. This means this model can be considered as a manybody quantum system.
A reasonable explanation is that the number of photons in the QRM also tends to infinity as $\omega/\omega_0$ tends to infinity, and then classical-quantum correspondence can be achieved and quantum chaotic signatures can be distinguished.
In this section, we numerically calculate the truncated photon number in QRM when the ratio of level-splitting $\omega$ to bosonic frequency $\omega_0$ grows to $\eta=\omega/\omega_0\rightarrow 18$.
In Fig. \ref{f2}, we exhibit the time evolution behavior of the average photon number $\langle \hat{a}^{\dagger}\hat{a}\rangle$ with different system photon number $N$ for the initial wave packet centred at point $C_1$.
It is shown that the dynamical evolution of $\langle \hat{a}^{\dagger}\hat{a}\rangle$ changes with the increase of the initial system photon number $N$.
However, when $N$ exceeds the critical value $N_c$($N_c \approx 140$), the evolutionary behaviors of $\langle \hat{a}^{\dagger}\hat{a}\rangle$ are consistent.
In other words, when the initial system photon number exceeds the critical value, i.e., $N>N_c$, the dynamic effects in the QRM no longer changes with the increase of photon number.
This indicates that the system photon number can be truncated at a value which greater than the critical value when the ratio of level-splitting $\omega$ to bosonic frequency $\omega_0$ grows to $18$.
Therefore, the large atom-light frequency ratios Rabi model can be considered as a semiclassical many-body quantum system when $\omega/\omega_0$ grows to large values, and then quantum chaos and classical-quantum correspondence can be studied in this system.
\begin{figure}[ht]
\begin{center}
\includegraphics[width=6cm]{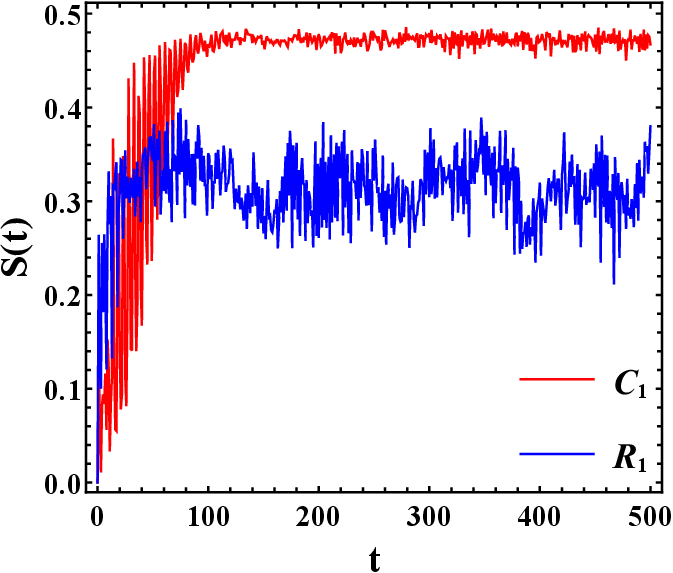}
\caption{The time evolution of linear entanglement entropy $S(t)$ for initial states centred at chaotic point $C_{1}$ and periodic point $R_{1}$ in Fig. \ref{f1}. Here, the system photon number is truncated at $150$.}
\label{f3}
\end{center}
\end{figure}

\section{LINEAR ENTANGLEMENT ENTROPY}\label{sec4}
\begin{figure}[ht]
\begin{center}
\includegraphics[width=6cm]{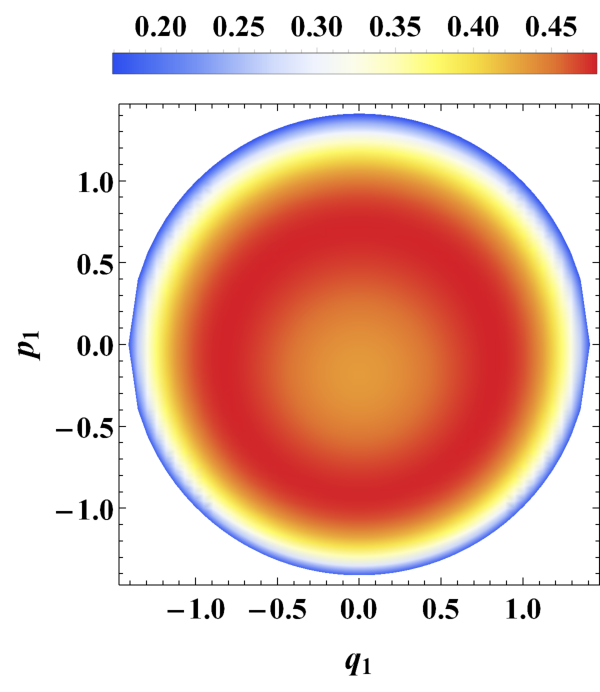}
\caption{The distribution of time-average entanglement entropy $S_{m}$ of Poincar\'{e} section in Fig. \ref{f1}. Here, the system photon number is truncated at $150$ and the integral interval is $t\in[0, 500]$.} \label{f4}
\end{center}
\end{figure}
In this section, we will investigate the quantum signatures of chaos in the QRM via linear entanglement entropy, which is an effective tool to explore chaos in quantum systems including the Dicke model~\cite{EE1,EE3,EECP2} and the kicked top model~\cite{EECP1}.
On the other hand, the buildup and decay of entanglement is closely linked with the collapse
and revival dynamics~\cite{husimiee}.
Therefore, it is necessary to re-examine the evolutional behaviors of linear entanglement entropy in chaotic and regular regions of the QRM with quantum collapse-revival effects.
The linear entanglement entropy is defined as
\begin{eqnarray}
S(t) &=& 1- \rm{Tr}_{1} \rho_{1}(t)^2.   \label{cha1}
\end{eqnarray}
with the reduced-density matrix
\begin{eqnarray}
\rho_{1} &=& \rm{Tr}_{2}|\psi(t)\rangle\langle\psi(t)|.   \label{cha2}
\end{eqnarray}
where $\rm{Tr}_{\mathrm{i}}$ is a trace over the $\mathrm{i}$ th subsystem($\mathrm{i}=1,2$) and the wave function $|\psi(t)\rangle$ is the quantum state of the full system.
The quantity $S(t)$ describes the degree of purity of the subsystems and the degree of decoherence.
In this paper, we take the time for linear entanglement entropy reaches saturated value as the critical time $t_c$. When $t>t_c$, we call it as long time.
In Fig. \ref{f3}, we exhibit the time evolution of linear entanglement entropy for different initial states. It is shown that the difference of linear entanglement entropy between chaotic and regular regions is very diminutive in short time.
\begin{figure*}[htb]
\centering
\epsfysize=7cm \epsfclipoff \fboxsep=0pt
\setlength{\unitlength}{1.cm}
\begin{picture}(10,5)(0,0)
\put(-4,-2.2){{\epsffile{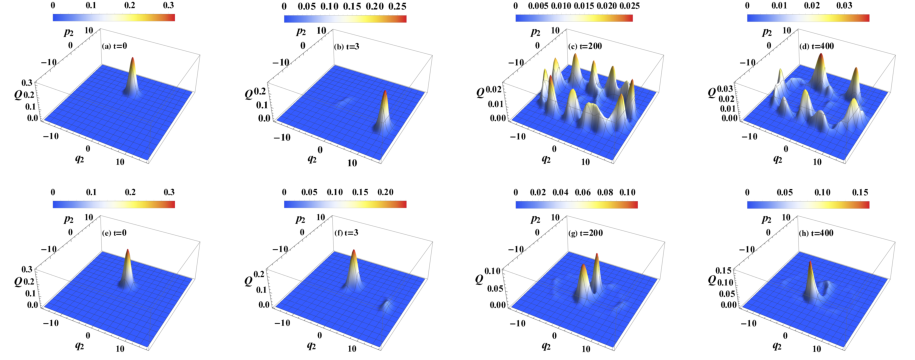}}}
\end{picture}
\vspace{2.3cm}
\caption{ The time evolution of the Husimi Q function.
The top and bottom panels denote respectively the case in which the initial coherent state centered at the points $C_1$ and $R_1$ in Fig. \ref{f1}. Here, the system photon number is truncated at $150$.}
\label{f5}
\end{figure*}
However, the saturated value of linear entanglement entropy for initial state located in the chaotic sea is significantly higher than that in the regular region.
Moreover, to avoid the randomness of the selection of initial state, we demonstrate the distribution of the time-average entanglement entropy $S_{m}=\frac{1}{T}\int_{t_{1}}^{t_{2}}S(t)\rm{d}t$ on the Poincar\'{e} section in Fig. \ref{f4}.
We clearly observe a significant dip in $S_{m}$ and more entanglement generation for initial states localized in chaotic sea compared to those in the regular regions.
This difference of linear entanglement entropy between chaotic and regular regions can be measured experimentally~\cite{EECP1}.
The correspondence between the phase space and the distribution of time-average linear entanglement entropy not only indicates that entanglement entropy can explore quantum chaos in the semiclassical QRM, but also indicates the classical-quantum correspondence can be achieved in QRM.

\section{HUSIMI Q FUNCTION}\label{sec5}

In the quantum domain, the quantum state is equivalent to its quasi-probability function in the phase space~\cite{husimi1}. Among the quasi-probability functions, the Husimi Q function allows one to visualize the dynamical evolution of quantum states in phase space. On the other hand, the collapse and revival of Husimi Q function have been interpreted theoretically ~\cite{husimicrthe2,husimicrthe3,husimicrthe4} and observed experimentally~\cite{husimicrexp1,husimicrexp2,husimicrexp3,husimicrexp4} in atom-field interaction systems. In this section, we focus on analyzing the time evolution difference of Husimi Q functions between chaotic and regular regions in the QRM.
For photon coherent state, the Husimi Q function is defined as
\begin{eqnarray}
Q(q_2,p_2) &=& \frac{1}{\pi} \langle q_{2},p_2|\rho_2|q_2,p_2\rangle,
\end{eqnarray}
where $|q_2,p_2\rangle$ is photon coherent state and $\rho_2$ is the reduced density matrix of the second subsystem.
In Fig. \ref{f5}, we present the time evolution of Husimi Q function in phase space.
Quantum collapse and revival indeed cause the Husimi Q function generate some effects which without classical dynamical counterparts such as splitting and merging, as shown in Fig. \ref{f5}(b) and Fig. \ref{f5}(f). The difference between Fig. \ref{f5}(b) and Fig. \ref{f5}(f) is extremely microscopic and cannot distinguish chaotic and regular orbits. However, at longer times $t>t_c$, the Husimi Q function with initial states located in chaotic sea are distributed in the outer part of the phase space, as shown in Fig. \ref{f5}(c)-Fig. \ref{f5}(d). This is different from that in the Dicke~\cite{NJC} and two-photon Dicke model~\cite{EECP2}, where the Husimi function in the chaotic region rapidly distributes throughout the whole phase space.
For initial states located in the regular regions, the Husimi Q function are concentrated in the vicinity of origin point, as shown in Fig. \ref{f5}(g)-Fig. \ref{f5}(h).
Obviously, the Husimi Q functions between chaotic and regular regions exhibit significant differences in long time.
Fig. \ref{f5} effectively indicates that the Husimi Q function can be considered as an effective tool for diagnosing chaos in the large atom-light frequency ratios Rabi model while quantum collapse-revival effects exist in this system.

\section{SPIN VARIANCE}\label{sec6}
\begin{figure}[ht]
\begin{center}
\includegraphics[width=6cm]{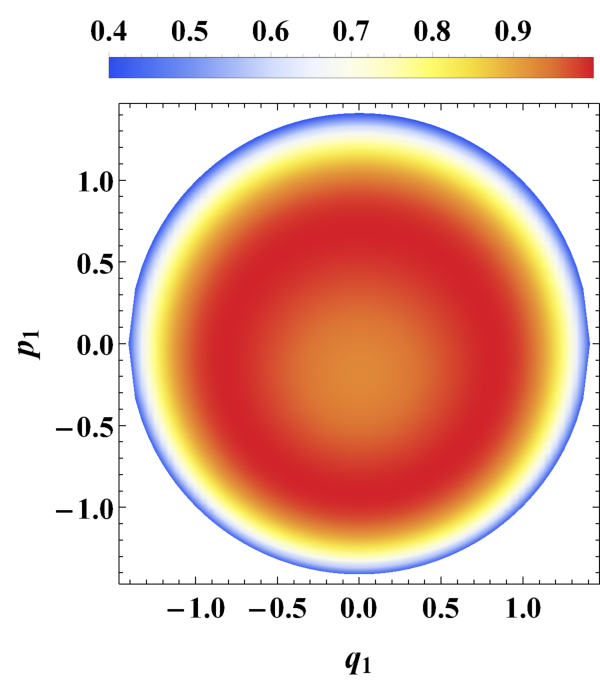}
\caption{The distribution of the time-average spin variance $ \Delta \sigma_z$ of Poincar\'{e} section in Fig. \ref{f1}. Here, the system photon number is truncated at $150$ and the time interval is $t\in[0, 500]$.} \label{f6}
\end{center}
\end{figure}
In this section, we will discuss the differences in spin variance between the chaotic sea and stable island in the Rabi model. For the Dicke model, in the spin limit, the spin variance value in chaotic regions are higher than that in regular regions~\cite{sl1}.
On the other hand, in the quantum Rabi model, the spin length $j=1/2$ is far from the spin limit $j\rightarrow\infty$ and the spin variance seems impossible to be as a tool for diagnosing quantum chaos. However, we still observed that the spin variance
$\Delta \sigma_z = \langle\sigma_{z}^2\rangle - \langle\sigma_{z}\rangle^2$ in chaotic regions are higher than that in regular regions, and the distribution of the long-time average $\Delta \sigma_z$ corresponds well with the semiclassical phase space in Fig. \ref{f1}, as shown in Fig. \ref{f6}.
This illustrates that in the bipartite interaction systems which far from the spin limit $j\rightarrow\infty$, the spin variance $\Delta \sigma_z$ still maintain its diagnostic function for quantum chaos in the large atom-light frequency ratios Rabi model.
The corresponding between the spin variance $\Delta \sigma_z$ and semiclassical phase space indicates that the atom in the Rabi model can exhibits some statistical characteristics of many-body systems. This means that classical-quantum correspondences in single atomic system can be achieved through interacting with light.

\section{SUMMARY}\label{sec7}
We have studied the long-time behaviors of linear entanglement entropy, Husimi Q function and spin variance in the large atom-light frequency ratios Rabi model.
It is shown that both the entanglement entropy and the Husimi Q function can still be used as tools to diagnose quantum chaos in the large atom-light frequency ratios Rabi model while quantum collapse-revival effects exist in
this system.
At longer times, the Husimi Q function for initial states located in the chaotic sea are distributed along the periphery of phase space. For initial states located in the regular regions, the Husimi Q function are concentrated in the vicinity of origin.
Moreover, we find that both the long-time average entanglement entropy and spin variance correspond well with the classical phase space.
Quantum collapse and revival indeed generate some quantum effects which without classical dynamics counterparts, such as the splitting and merging of wave packets, but it does not render the correspondence principle ineffective at longer times. Our conclusion not only suggests that the correspondence principle does not invalidated by quantum collapse and revival effects in the QRM, but also suggests that the multi-body statistical characteristics in a single atomic system can be achieved through interacting with light.

\section{Acknowledgments}
This work was supported by the National Natural Science Foundation of China under Grant No.12275078,
11875026, 12035005, 2020YFC2201400, and Science Foundation of Hengyang Normal University of China under Contract No. 2020QD24. This work is also sponsored by the innovative research group of Hunan Province under Grant No. 2024JJ1006.

\end{document}